\documentclass[shortnote,twocolumn]{jpsj2}
\usepackage{graphicx}

\def\runtitle{Thermoelectric properties of disordered systems}
\def\runauthor{R.A.~\textsc{R\"{o}mer}, A.~\textsc{MacKinnon} \&
C.~\textsc{Villagonzalo}}

\title{%
Thermoelectric properties of disordered systems}

\author{%
  Rudolf A. \textsc{R\"{o}mer}$^{1}$\thanks{Permanent address: Physics
    Department, University of Warwick, Coventry CV4 7AL, UK,
    Email: r.roemer@warwick.ac.uk}, Angus \textsc{MacKinnon}$^{2,3}$, and Cristine \textsc{Villagonzalo}$^4$}

\inst{%
$^1$Institut f\"{u}r Physik, Technische Universit\"{a}t, D-09107 Chemnitz,
  Germany\\
$^2$Blackett Laboratory, Imperial College, London SW7 2BW, UK\\
$^3$Cavendish Laboratory, Madingly Rd, Cambridge CB3 0HE, UK\\
$^4$National Institute of Physics, University of the Philippines, Quezon City, Philippines
}

\recdate{$Revision: 1.10 $, compiled \today}

\abst
{
}
\kword{%
thermopower,localization, recursive Green function method
}

\begin{document}
\sloppy
\maketitle

%%%%%%%%%%%%%%%%%%%%%%%%%%%%%%%%%%%%%%%%%%%%%%%%%%%%%%
\section{Introduction}
%%%%%%%%%%%%%%%%%%%%%%%%%%%%%%%%%%%%%%%%%%%%%%%%%%%%%%

The electronic properties of disordered systems have been the subject
of intense study for several decades.  Thermoelectric properties, such
as thermopower and thermal conductivity, have been relatively
neglected. A long standing problem is represented by the sign of the
thermoelectric power \cite{VilRS99a}.  In crystalline semiconductors
this is related to the sign of the majority carriers, but in
non-crystalline systems it is commonly observed to change sign at low
temperatures.  In spite of its apparent universality this change has
been interpreted in a variety of ways in different systems. We have
developed a Green's function recursion algorithm \cite{Mac85} based
on the Chester-Thellung-Kubo-Greenwood formula \cite{CheT61,Gre58,Kub57} for
calculating the kinetic coefficients $L_{ij}$ on long strips or bars.
From these we can deduce the electrical conductivity $\sigma$, the
Seebeck and Peltier coefficients $S$ \& $\Pi$ and the thermal
conductivity $\kappa$, as well as the Lorenz number $L_0$. We present
initial results for 1D systems. In 1D we observe a Lorentzian-like
distribution for the thermopower which is modified by the presence of
inelastic scattering. This could give rise to non-negligible quantum
fluctuations in macroscopic systems at low temperatures.

%%%%%%%%%%%%%%%%%%%%%%%%%%%%%%%%%%%%%%%%%%%%%%%%%%%%%%
%\section{Linear response formulation}
%%%%%%%%%%%%%%%%%%%%%%%%%%%%%%%%%%%%%%%%%%%%%%%%%%%%%%

Within the linear response approach, the responses of a system to an
external electric field $\mathbf{E}$ and a temperature gradient
$\mathbf{\nabla}T$ up to linear order \cite{Cal85} are
\begin{eqnarray}
  \langle\mathbf{j}\rangle 
& = & |e|^{-1}\left(|e|L_{11}
    \mathbf{E}-L_{12}T^{-1}\mathbf{\nabla}T\right)\nonumber \\
& = &
  \sigma\mathbf{E}-\sigma{S}\nabla{T} \label{eq:ecurrent}\\
   \label{p:lij}
%\end{eqnarray}
%-------------****---------------
%and
%-------------****---------------
%\begin{eqnarray}
  \langle\mathbf{j}_{q}\rangle 
& = & e^{-2}\left(|e|L_{21}
    \mathbf{E}-L_{22}T^{-1}\mathbf{\nabla}T\right)\nonumber \\
& = &
    ST\langle\mathbf{j}\rangle
  - K\nabla{T}
%    S\langle\mathbf{j}\rangle
%  -\frac{K\nabla{T}}{T}
  \label{eq:hcurrent}.
\end{eqnarray}
Here, $\langle\mathbf{j}\rangle$ and $\langle\mathbf{j}_{q}\rangle$
denote charge and heat currents, $L_{ij}$, $i,j= 1,2$ are the kinetic
coefficients and $e$, $T$, $\sigma$, $S$, and $K$ are electron charge,
temperature, conductivity, thermopower and heat conductivity,
respectively. Other commonly considered thermoelectric transport
coefficients are the Peltier coefficient, $\Pi$, and the Lorenz number,
$L_0$
%-------------****---------------
\begin{eqnarray}
  \Pi &=& \frac{L_{12}}{|e|L_{11}} = S T \label{eq:pel2}\\
  L_{0}&=&\frac{L_{22}L_{11}-L_{21}L_{12}}{(k_{\mbox{\scriptsize B}}TL_{11})^2} 
  = \frac{e^2}{k_{\mbox{\scriptsize B}}^2}\frac{K}{\sigma{T}}\;, \label{eq:lo1}
\end{eqnarray}
with $k_{\rm B}$ the Boltzmann constant.  Calculations of the
thermoelectric transport properties that have been done previously are
based either on the CTKG formulation\cite{CheT61,Gre58,Kub57} or other
perturbative methods\cite{CasCGS88,SivI86}.  These methods are
macroscopic in nature and rely on the explicit, phenomenological form
$\sigma(E_{\rm F},T)$. A microscopic approach to determine the
thermoelectric transport properties without assuming explicitly a
functional form of $\sigma$ would be quite instructive.  To our
knowledge, there has only been Ref.\ \cite{LanSB98} which reported on
the distribution of $S$ in a 1D disordered wire in the localized
regime. The authors were able to analytically and numerically show
that in this case the distribution of $S$ is Lorentzian.  So far there
has been no other microscopic calculation of $S$, $\Pi$, $K$ or $L_0$
in disordered systems near the MIT.

%%%%%%%%%%%%%%%%%%%%%%%%%%%%%%%%%%%%%%%%%%%%%%%%%%%%%%
\section{The recursive Green function method}
%%%%%%%%%%%%%%%%%%%%%%%%%%%%%%%%%%%%%%%%%%%%%%%%%%%%%%

This method was used previously to calculate
the dc and ac conductivity tensors and the density of states of a
$d$-dimensional disordered system \cite{Mac80,Mac85} with and without
a magnetic field.  Here we will show briefly how the method can be
adopted to calculate all kinetic coefficients and thus the other
transport properties.

The 1D Hamiltonian can be written as
%-------------****---------------
\begin{equation}
%  H=\sum_{j=1}^{N} \epsilon_{j}|j\rangle\langle{j}|+ 
%    \left(|j\rangle\langle{j+1}|+|j+1\rangle\langle{j}|\right)
H=\sum_{j=1}^N \epsilon_j c^\dagger_j c_j + c^\dagger_j c_{j+1} + c^\dagger_{j+1}c_j
  \,,
\label{eq:hamilton2}
\end{equation}
%-------------****---------------
where the disorder terms $\epsilon_{j}\in[-W/2,W/2]$ are randomly
distributed.  In adding one more site to the chain, the Hamiltonian
can then be considered as an unperturbed Hamiltonian of the first $N-1$
sites and the decoupled $N^{\mbox{\scriptsize\rm th}}$ site plus a perturbation
\cite{Mac80}.  Equation (\ref{eq:hamilton2}) then simplifies to
%-------------****---------------
\begin{equation}
\!H^{(N)}\!=\!\left[H^{(N-1)}\! + \epsilon_{N} c^\dagger_{N} c_{N}\right]
+\left[c^\dagger_{N-1} c_{N} + c^\dagger_{N} c_{N-1}\right]
\label{eq:hamilton5}
\end{equation}
%-------------****---------------
The last part in the Hamiltonian, $H'$, is the coupling between the
$(N-1)^{\mbox{\scriptsize\rm th}}$ site and the $N^{\mbox{\scriptsize\rm
th}}$ site.  This coupling can be treated iteratively using the
one--particle Green function $G^{\pm}(\mathrm{z}^{\pm})$
%-------------****---------------
%\begin{equation}
%  G^{\pm}_{jk} \equiv \langle j \mid G^{\pm}(z)\mid{k}
%  \rangle=\sum_{i}\frac{\langle {j}\mid{i}\rangle\langle{i}\mid{k}
%    \rangle}{\mathrm{z}^{\pm}-E_i}\,
%\label{eq:grn4}
%\end{equation}
%-------------****---------------
as
%-------------****---------------
\begin{equation}
  G^{(N)}=G_0^{(N-1)}+G_0^{(N-1)}H'G^{(N)}\,
\label{eq:grn2}
\end{equation}
where 
$\mathrm{z}^{\pm} = E \pm i\xi \label{p:complexE}$, where $E$ is the energy
and $\xi$ may be interpreted as an inverse lifetime, \mbox{e.g.} due to
inelastic scattering.

%%%%%%%%%%%%%%%%%%%%%%%%%%%%%%%%%%%%%%%%%%%%%%%%%%%%%%
%\section{Results for $L_{12}$}
\section{Results and Conclusions}
%%%%%%%%%%%%%%%%%%%%%%%%%%%%%%%%%%%%%%%%%%%%%%%%%%%%%%

\begin{figure}[h]
\centerline{\includegraphics[scale=0.4]{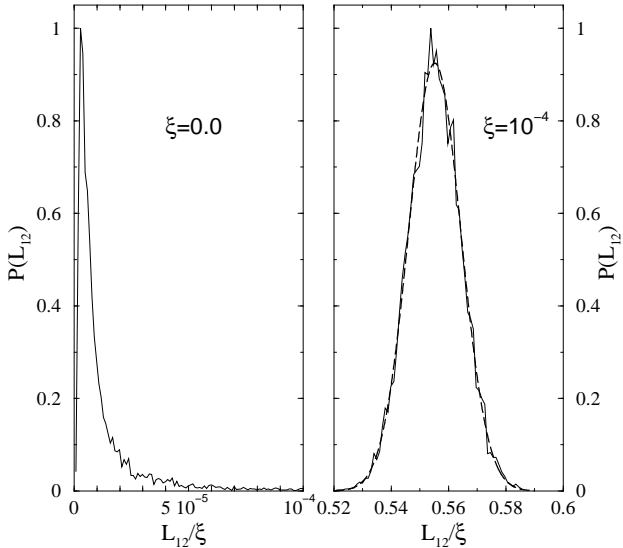}}
%\centerline{\psfig{figure=fig-fss-change.eps,width=\textwidth}}
\caption{\label{fig-distP-L12}
  Distribution function $P(L_{12})$ for zero and finite value of $\xi$
  at $E=\mu=-1$, $W=2$ and $T=0$. Data is computed from 10000 samples
  and length $Na= 3\times 10^{6}a$. $P$ is normalized to $1$ at the
  maximum. The dashed line correspond to a Gaussian (right) fit.  }
\end{figure}

For brevity, we shall restrict ourselves here to studying the behavior
of $S$ ($L_{12}$). In complete analogy to the formulation
for $\sigma$ ($=L_{11}$)\cite{Mac80,Mac85}, we can compute $L_{12}(T=0)$ 
using the recursive equations
%-------------****---------------
\newcommand{\negspace}{\hspace{-8pt}}
\begin{eqnarray}
L^{(N)}_{12}\negspace&=&\negspace(E-\mu)\sigma^{(N)}
-\xi\frac{e^2}{\pi \hbar N a}
     \mathrm{Im}f^+_N~, \label{eq:l12_site}\nonumber \\
h^+_N\negspace&=&\negspace g_N^{+2} \left[ (x_{N}\!-\! x_{N-1})^2 g_{N-1}^+ +
    h^+_{N-1} \right]~,\label{eq:hN} \\
f^+_N\negspace&=&\negspace f^+_{N-1}\!-2g_{N}^+ \left[(x_{N}\!-\! x_{N-1})^2 g_{N-1}^+
+ h^+_{N-1}\right] \label{eq:fN}\nonumber .
\end{eqnarray}
%-------------****---------------
The initial condition is $f_0^{+} = h_0^{+} = 0$.  From Eq.\ 
(\ref{eq:pel2}) we see that
%-------------****---------------
\begin{equation}
  \Pi^{(N)}=\frac{1}{|e|} \left\{ (E-\mu) - \frac{\xi}{4}
    \frac{ \mathrm{Im}f^+_{N} }{s_{N}} \right\} 
\propto L_{12}^{(N)}\,.
\label{eq:PiN}
\end{equation}
%-------------****---------------
where $s_{N}= {\pi \hbar Na}\sigma^{(N)} / {4 e^2} $. The
distribution of $L_{12}$ (in units of $e^2/\pi\hbar a$) is shown in Fig.\ \ref{fig-distP-L12}.

\begin{figure}%[h]
\vspace{-0.5cm}
\centerline{\includegraphics[scale=0.4]{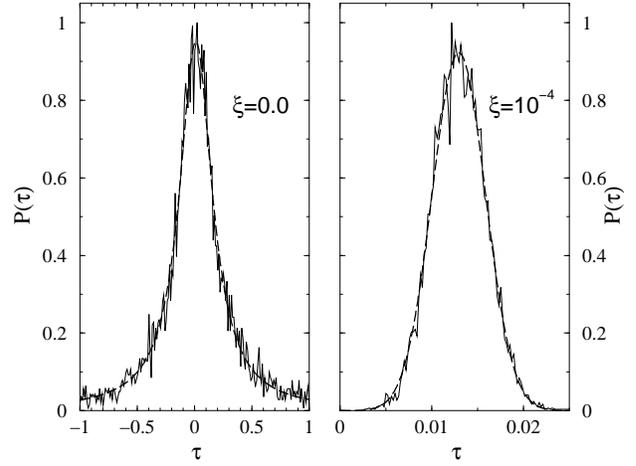}}
%\centerline{\psfig{figure=fig-fss-change.eps,width=\textwidth}}
\caption{\label{fig-distP-DPC}
  Distribution function $P(\tau)$ for zero and finite value of $\xi$ and
  the same parameters as Fig.~\ref{fig-distP-L12}. The dashed
lines correspond to a Lorentzian (left) and a Gaussian (right) fit.}
\end{figure}

A further interesting quantity is the {\em dimensionless thermopower}
$\tau$ defined via
%-------------****---------------
\begin{equation}
\tau = \left.\frac{\Delta}{2\pi\mathbf{g}(E)}
\frac{d\mathbf{g}(E)}{dE}\right|_{E=E_{\mbox{\scriptsize F}}}
=
\frac{\Delta}{2\pi}\frac{d\ }{dE}
   \ln \mid{G}^+_{1N}(E) {\mid}^2\;,
\label{eq:dimS1}
\end{equation}
%-------------****---------------
with $\Delta$ the mean level spacing near $E_{\mbox{\scriptsize F}}$
and $\mathbf{g}$ the conductance.  In the limit of weak scattering the
dominant effect of the disorder is simply to introduce a finite mean
free path at energies well inside the band. This implies that on the
average (\ref{eq:dimS1}) can be
approximated by
%-------------****---------------
$%\begin{equation}
  \tau= - \frac{\Delta}{\pi} \mbox{Tr} \left[ \mathrm{Re}\,G^+(E)
  \right]
$. %\end{equation}
%-------------****--------------
The full distribution can be shown to have a simple Lorentzian form \cite{LanSB98}
%-------------****---------------
%\begin{equation}
%  P(\tau)
%  =\frac{1}{\pi}\frac{1}{ 1 + \tau^2 }
%\label{eq:Sdistrib}
%\end{equation}
%-------------****---------------
which can be compared with the results of the recursive approach as shown in Fig.~\ref{fig-distP-DPC}. \\
\\

%%%%%%%%%%%%%%%%%%%%%%%%%%%%%%%%%%%%%%%%%%%%%%%%%%%%%%
%\section{Summary}
%%%%%%%%%%%%%%%%%%%%%%%%%%%%%%%%%%%%%%%%%%%%%%%%%%%%%%

Using $O(N)$ arithmetic operations, where $N$ is the number of sites
in the system, we can compute iteratively all kinetic transport
coefficients $L_{ij}$.  The maximum value of $N$ is mainly limited by
the computer time available and the intended precision of the results.
In this short note, we have considered only the transport properties
of a chain initially having $N$ sites of a 1D tight-binding model with
nearest neighbor hopping in the limit $T=0$.  It is straightforward to
extend the computations to higher dimensions \cite{KraS96} and results
will be presented elsewhere. Detailed derivations of relevant
equations and recursion relations are given elsewhere \cite{Vil01}.

We have shown that the distribution of the dimensionless thermopower
$\tau$ is indeed Lorentzian as predicted\cite{LanSB98} but that this
changes to Gaussian on introduction of a finite lifetime.  Similarly
the distribution of $L_{12}$ is Lorentzian--like (albeit asymmetric)
for $\xi\rightarrow 0$ but also becomes Gaussian for finite $\xi$.

\end{document}